\documentclass[12pt,a4]{article}

\usepackage{a4}
\usepackage{amstext}
\usepackage{amsfonts}
\usepackage{amssymb}
\usepackage{color}
\usepackage{epsfig}
\usepackage{verbatim}

\newcommand{\gtapprox}{\raisebox{-0.5ex}{$\,\stackrel{>}{\scriptstyle\sim}\,$}}
\newcommand{\ltapprox}{\raisebox{-0.5ex}{$\,\stackrel{<}{\scriptstyle\sim}\,$}}

\begin{document}


\begin{center}

{\Large \bf An introduction to lattice hadron spectroscopy for}

\vspace{-0.1cm}
{\Large \bf students without quantum field theoretical}

\vspace{-0.1cm}
{\Large \bf background}

\vspace{0.8cm}

{\large \bf HGS-HIRe Lecture Week on Hadron Physics -- July 2013 -- Laubach, Germany}

\vspace{0.8cm}

\textbf{Marc Wagner} \\
Goethe-Universit\"at Frankfurt am Main, Institut f\"ur Theoretische Physik, \\ 
Max-von-Laue-Stra{\ss}e 1, D-60438 Frankfurt am Main, Germany

\vspace{0.8cm}

lecture notes written by

\vspace{0.2cm}

\textbf{Stefan Diehl, Till Kuske} \\
Justus Liebig-Universit\"at Giessen, II.~Physikalisches Institut, \\ Heinrich-Buff-Ring 16, D-35392 Giessen, Germany

\vspace{0.2cm}

\textbf{Johannes Weber} \\
Johannes Gutenberg-Universit\"at Mainz, Institut f\"ur Kernphysik, Johann-Joachim-Becher-Weg 45, D-55099 Mainz, Germany

\vspace{0.8cm}

October 7, 2013

\end{center}

\vspace{0.4cm}

\begin{tabular*}{16cm}{l@{\extracolsep{\fill}}r} \hline \end{tabular*}

\vspace{-0.4cm}
\begin{center} \textbf{Abstract} \end{center}
\vspace{-0.4cm}

The intention of these lecture notes is to outline the basics of lattice hadron spectroscopy to students from other fields of physics, e.g.\ from experimental particle physics, who do not necessarily have a background in quantum field theory. After a brief motivation and discussion of QCD, it is explained, how QCD can in principle be solved numerically using lattice QCD. The main part of these lecture notes is concerned with quantum numbers of hadrons, corresponding hadron creation operators, and how the mass of a hadron can be determined from a temporal correlation function of such operators. Finally, three recent lattice hadron spectroscopy examples from the literature are discussed on an elementary level.

\begin{tabular*}{16cm}{l@{\extracolsep{\fill}}r} \hline \end{tabular*}

\thispagestyle{empty}


\newpage

\setcounter{page}{1}

\section{Introduction}

These lectures on lattice QCD and the computation of hadron masses were given at the HGS HIRe Lecture Week on Hadron Physics 2013. They are primarily aimed at non-experts from other fields, e.g.\ from experimental particle physics, possibly without any knowledge of quantum field theory. The goal of these lectures is to provide insight into the methods of lattice QCD sufficient to grasp the main concepts of papers and presentations on the subject.

In particular, the lectures outline, how masses of hadrons -- i.e.\ mesons ($\bar q q$ states like $\pi$, $K$, $D$, $\rho$, ...) and baryons ($q q q$ states like $p$, $n$, ...) -- can be calculated numerically from first principles. ``First principles'' implies that the calculation is exclusively based on the QCD action without any simplifying assumptions or approximations. The input parameters are the quark masses ($m_u$, $m_d$, $m_s$, ...).

There are many reasons to be interested in studying the strong interactions by means of lattice QCD. For example trying to reproduce the experimentally observed properties of hadrons by a first principles computation allows to verify QCD and the Standard Model as the correct theory of particle physics up to the given numerical and experimental precision or, similarly, to search for new physics (e.g.\ lattice computations of the hadronic contribution to the muonic anomalous moment [$g-2$] are indispensable in order to further reduce the error of the theoretical prediction). Another benefit of lattice QCD is its ability to predict hadronic states, which have not been discovered experimentally yet. Such results can provide useful input both for phenomenologists and for future experiments. Moreover, lattice QCD can be used to determine quantum numbers and the structure of experimentally less well-established states (states ``omitted from summary table'' by PDG \cite{PDG}).

Typical properties of hadrons (e.g.\ masses, structure functions, decay constants, ...) are dictated by QCD, which is briefly discussed in the next section. On the other hand, QED and the weak interactions cause only small corrections, which can often be neglected or estimated by perturbation theory or model calculations. Finally, gravitation is irrelevant for hadronic properties.


\section{QCD}

QCD is the quantized field theory of quarks and gluons. The fields of the theory are divided into matter and force fields. The matter fields consist of six quark fields,
\begin{eqnarray}
\label{EQN706} \psi^{(f)}(x) \ \ \equiv \ \ \psi^{(f)}(\mathbf{r},t) \quad , \quad f \ \ \in \ \ \{ u,d,s,c,t,b \} ,
\end{eqnarray}
which are treated in QCD as exact copies, which differ only in terms of their masses. The force field is a single massless gluon field,
\begin{eqnarray}
\label{EQN707} A_\mu(x) \ \ \equiv \ \ A_\mu(\mathbf{r},t) .
\end{eqnarray}


\subsection{Why do we need a quantized field theory?}

One could phrase this question also in other words: ``\textit{Why is it insufficient to use quantum mechanics in order to describe the strong interactions?}''

In particle physics creation and annihilation of particles is important (high energy collider experiments are an obvious example). Quantum mechanics, however, describes each particle by a wave function. The absolute square of a wave function $\psi(\mathbf{r})$ is the probability density to find the particle at $\mathbf{r}$. Conservation of probability, which follows directly from Schr\"odinger's equation, does neither allow for the annihilation of an existing particle nor the creation of a new particle. Hence, creation and annihilation processes cannot be described by quantum mechanics.

Despite $E = m c^2$ dictating the relation between energy and mass in particle creation or annihilation, also relativistic quantum mechanics leaves the problem of describing creation and annihilation processes unresolved (the absolute square of a wave function is still interpreted as a probability density).

Field theory, on the other hand, allows for the creation and annihilation of particles. As an analogy, the field can be visualized as a fine 3-dimensional network of springs (``a 3-dimensional spider web''). If these springs are at rest at their energetic minimum, i.e.\ if there are no oscillations, an experimentalist would not observe a particle. If the springs oscillate in a certain spatial region, the field carries localized energy. An experimentalist would interpret this energy as one or more particles in that region. In the case of QCD a quark field and the gluon field can be visualized as two independent networks of springs, which are connected by additional springs. Clearly, energy associated with oscillations of the quark field can now be transferred to the gluon field and vice versa. This is, how the mechanism of particle creation and annihilation works in field theory: e.g.\
\begin{itemize}
\item a(n) (anti)quark emits a gluon or

\item a(n) (anti)quark absorbs a gluon or

\item a quark and a antiquark annihilate, resulting in a gluon or

\item ...
\end{itemize}

One could now ask: ``\textit{Why does it have to be a quantized field theory? Why is it insufficient to just use a classical field theory as e.g.\ in electrodynamics?}''

In a classical field theory field excitations can carry any amount of energy. Thus, there is no integer quantity, which can be interpreted as the number of particles. In typical problems from classical electrodynamics, where millions of photons are present, this is not a problem and classical field theory works rather well. In contrast to that, in a quantum field theory field excitations are quantized, i.e.\ corresponding energies are discretized (similarly as for the quantized harmonic oscillator). There exists an integer quantity, which corresponds to the number of particles. In QCD problems, where the number of particles is typically rather small (e.g.\ three quarks, when studying a baryon), this is essential for an accurate theoretical description.

In the following it is not assumed that the reader possesses a working knowledge of quantum field theory. Aspects of quantum field theory will be avoided, whenever possible. If not possible, the necessary concepts will be vaguely illustrated. The goal of these notes is to outline the basic ideas behind the computation of hadron masses using lattice QCD. However, if one is interested in understanding also the details, there is no other way than to study quantum field theory (cf.\ e.g.\ the textbooks \cite{Ryde,Sred,Magg,PeSc}).


\subsection{The QCD action}

The fields (\ref{EQN706}) and (\ref{EQN707}) appearing in the QCD action have several components, which are labeled by various indices.

The quark fields, $\psi^{a,(f)}_{A}$, have a color index $a = 1,\ldots,3$ (quarks carry color charge; in contrast to electrical charge there are three types of color charge, red, green and blue; the expert would say ``QCD is an SU(3) gauge theory''). The three color components can be seen as the three entries of a column vector. Since quarks are spin-$1/2$ particles, there is also a Dirac or spin index $A = 1,\ldots,4$ (the four degrees of freedom correspond to spin up/spin down and to particle/antiparticle). Finally, quarks have a flavor index $(f) = 1,\ldots,N_f$ (in full QCD $N_f = 6$, i.e.\ there are six quark flavors, $u$, $d$, $s$, $c$, $b$ and $t$ quarks, which differ in mass, $m_u = 2.3 \, \textrm{MeV}$, $m_d = 4.8 \, \textrm{MeV}$, $m_s = 95 \, \textrm{MeV}$, $m_c = 1.28 \, \textrm{GeV}$, $m_b \approx 4 \,\textrm{GeV}$, $m_t \approx 170 \, \textrm{GeV}$\footnote{Throughout these lecture notes natural units are used, i.e.\ $\hbar = c = 1$; e.g.\ $1 \, \textrm{MeV} \equiv 1.78 \times 10^{-30} \, \textrm{kg}$.} \cite{PDG}).

The gluon field $A_\mu^a$ also has a color index $a = 1,\ldots,8$. Often it is convenient to write the gluon field as a matrix, $A_\mu = A_\mu^a \lambda^a / 2$, where $\lambda^a$ denote the eight $3 \times 3$ Gell-Mann matrices, e.g.\
\begin{eqnarray}
\lambda^1 \ \ = \ \ \left(\begin{array}{ccc}
0 & +1 & 0 \\
+1 & 0 & 0 \\
0 & 0 & 0
\end{array}\right) \quad , \quad \lambda^2 \ \ = \ \ \left(\begin{array}{ccc}
0 & -i & 0 \\
+i & 0 & 0 \\
0 & 0 & 0
\end{array}\right) \quad , \quad \ldots
\end{eqnarray}
The three rows and columns of the Gell-Mann matrices correspond to the quark colors red, green and blue. For example gluons, which are excitations of the field component $A_\mu^1$, mediate forces between red and green quarks. Furthermore, the gluon field has a Lorentz index or spacetime index $\mu = 0,\ldots,3$, since gluons are spin-$1$ particles.

The quark part of the QCD action depends on the quark fields as well as on the gluon field:
\begin{eqnarray}
\nonumber & & \hspace{-0.7cm} S_\textrm{quark}[\psi,\bar{\psi},A] \ \ = \ \ \int d^4x \, \sum_f \bar{\psi}^{(f)} \Big(i \gamma^\mu D_\mu - m^{(f)}\Big) \psi^{(f)} \ \ = \\
\label{EQN755} & & = \ \ \int d^4x \, \sum_f \bar{\psi}_A^{a,(f)} \Big(i \gamma^\mu_{A B} \Big(\delta^{a b} \partial_\mu - i g A_\mu^c \lambda^{c,a b}/2\Big) - \delta^{a b} \delta_{A B} m^{(f)}\Big) \psi_B^{b,(f)} .
\end{eqnarray}
$\bar{\psi} = \psi^\dagger \gamma^0$, $D_\mu = \partial_\mu - i g A_\mu$ denotes the so-called covariant derivative, where $g$ is the QCD coupling constant (similar to the electrical charge in electrodynamics) and $\gamma_\mu$ are the $4 \times 4$ Dirac matrices, which can be chosen according to
\begin{eqnarray}
\gamma^0 \ \ = \ \ \left(\begin{array}{cc} +1 & 0 \\ 0 & -1 \end{array}\right) \quad , \quad \gamma^j \ \ = \ \ \left(\begin{array}{cc} 0 & +\sigma_j \\ -\sigma_j & 0 \end{array}\right) ,
\end{eqnarray}
where $\sigma_j$ are the Pauli matrices,
\begin{eqnarray}
\sigma_1 \ \ = \ \ \left(\begin{array}{cc} 0 & +1 \\ +1 & 0 \end{array}\right) \quad , \quad \sigma_2 \ \ = \ \ \left(\begin{array}{cc} 0 & -i \\ +i & 0 \end{array}\right) \quad , \quad \sigma_3 \ \ = \ \ \left(\begin{array}{cc} +1 & 0 \\ 0 & -1 \end{array}\right) .
\end{eqnarray}
Note the similarity of (\ref{EQN755}) to the Dirac equation, $(i \gamma^\mu \partial_\mu - m) \psi = 0$, which is typically discussed in standard lectures on quantum mechanics.

The gluon part of the QCD action depends only on the gluon field:
\begin{eqnarray}
\label{EQN756} S_\textrm{gluon}[A] \ \ = \ \ -\frac{1}{4} \int d^4x \, F^{\mu \nu,a} F_{\mu \nu}^a \quad , \quad F_{\mu \nu}^a \ \ = \ \ \partial_\mu A_\nu^a - \partial_\nu A_\mu^a + g f^{a b c} A_\mu^b A_\nu^c
\end{eqnarray}
or equivalently
\begin{eqnarray}
\nonumber & & \hspace{-0.7cm} S_\textrm{gluon}[A] \ \ = \ \ -\frac{1}{2} \int d^4x \, \textrm{Tr}\Big(F^{\mu \nu} F_{\mu \nu}\Big) \quad , \\
 & & \hspace{0.675cm} \quad F_{\mu \nu} \ \ = \ \ F_{\mu \nu}^a \frac{\lambda^a}{2} \ \ = \ \ \partial_\mu A_\nu - \partial_\nu A_\mu - i g [A_\mu , A_\nu] .
\end{eqnarray}
Here, the $f^{a b c}$ are the totally antisymmetric structure constants of SU(3), $[\lambda^a/2 , \lambda^b/2] = i f^{a b c} \lambda^c / 2$.

Note the similarity of (\ref{EQN756}) to the action of electrodynamics, $S = -(1/4) \int d^4x \, F^{\mu \nu} F_{\mu \nu}$, $F_{\mu \nu} = \partial_\mu A_\nu - \partial_\nu A_\mu$. The term $g f^{a b c} A_\mu^b A_\nu^c$ in (\ref{EQN756}), which has no counterpart in electrodynamics, is responsible for interactions between gluons.

The QCD action is the sum of the quark action and the gluon action:
\begin{eqnarray}
S_\textrm{QCD}[\psi,\bar{\psi},A] \ \ = \ \ S_\textrm{quark}[\psi,\bar{\psi},A] + S_\textrm{gluon}[A] .
\end{eqnarray}


\subsection{Quantization of QCD}

In quantum mechanics the coordinate $x$ and the momentum $p$ of a particle are replaced by corresponding operators $\hat{x}$ and $\hat{p}$. Imposing the commutation relation $[x,p] = i$ is sufficient, to generate quantum effects. When quantizing fields, e.g.\ the quark and the gluon fields in QCD, the procedure is the same as in quantum mechanics. $\psi^{(f)}$ and $A_{\mu}$ are replaced by operators $\hat{\psi}^{(f)}$ and  $\hat{A}_{\mu}$, which obey appropriate commutation relations.

There are several mathematically equivalent approaches to calculate quantum mechanical or quantum field theoretical expectation values:
\begin{itemize}
\item[(1)] \textbf{Schr\"odinger's equation, wave functions} \\
This approach is widely used in quantum mechanics and should be well known from standard lectures. As an example, the ground state mean square displacement of a particle is given by
\begin{eqnarray}
\langle 0 | \hat{x}^2 | 0 \rangle \ \ = \ \ \int dx \, \psi_0^\ast(x) x^2 \psi_0(x) ,
\end{eqnarray}
where $\psi_0(x)$ is the ground state wave function, i.e.\ that solution of Schr\"odinger's equation with the lowest energy eigenvalue.

Generalizing this approach to quantum field theory is, however, not very practical (in contrast to mechanics there are infinitely many degrees of freedom in field theory). Therefore, it is rarely used in quantum field theory or QCD.

\item[(2)] \textbf{Creation and annihilation operators} \\
In quantum mechanics creation and annihilation operators are usually introduced in the context of the harmonic oscillator:
\begin{eqnarray}
\hat{a} \ \ = \ \ \sqrt{\frac{m \omega}{2}} \hat{x} + i \sqrt{\frac{1}{2 m \omega}} \hat{p} \quad \rightarrow \quad \hat{a}^\dagger \ \ = \ \ \sqrt{\frac{m \omega}{2}} \hat{x} - i \sqrt{\frac{1}{2 m \omega}} \hat{p} .
\end{eqnarray}
One can show that applying a creation/an annihilation operator to an energy eigenstate $| n \rangle$ yields $| n + 1 \rangle$/$| n - 1 \rangle$, i.e.\ creates/annihilates one quantum of energy:
\begin{eqnarray}
\hat{a}^\dagger | n \rangle \ \ = \ \ \sqrt{n+1} | n+1 \rangle \quad , \quad \hat{a} | n \rangle \ \ = \ \ \sqrt{n} | n-1 \rangle .
\end{eqnarray}
If a theory does not have quadratic form, but is similar to a quadratic theory, the method of creation and annihilation operators can still be applied. Perturbation theory is then used, to deal with non-quadratic terms. Specializing the above example to the case of the harmonic potential $V(x) = m \omega^2 x^2 / 2$ the mean square displacement can be expressed in terms of creation and annihilation operators:
\begin{eqnarray}
\langle 0 | \hat{x}^2 | 0 \rangle \ \ = \ \ \langle 0 | \sqrt{\frac{1}{2 m \omega}} \Big(\hat{a} + \hat{a}^\dagger\Big) \sqrt{\frac{1}{2 m \omega}} \Big(\hat{a} + \hat{a}^\dagger\Big) | 0 \rangle \ \ = \ \ \ldots \ \ = \ \ \frac{1}{2 m \omega} .
\end{eqnarray}

The method of creation and annihilation operators is widely used in perturbative quantum field theory. However, hadron masses cannot be calculated perturbatively. This is so, because the QCD coupling constant $g$, the expansion parameter of perturbative QCD, is only small for large quark and gluon momenta, while hadrons also contain quarks and gluons with small momenta.

\item[(3)] \textbf{Path integrals} \\
A less common approach to calculate ground state expectation values in quantum mechanics are path integrals (cf.\ e.g.\ \cite{Ryde,Rothe} for a detailed derivation). Using the path integral approach the above example of the mean square displacement reads
\begin{eqnarray}
\label{EQN891} \langle 0 | \hat{x}^2 | 0 \rangle \ \ = \ \ \frac{1}{Z} \int Dx \, x(0)^2 e^{i S[x]} \quad , \quad Z \ \ = \ \ \int Dx \, e^{i S[x]} .
\end{eqnarray}
$\int Dx = \prod_t dx(t)$ is the integral over all possible paths $x(t)$ from $t = -\infty$ to $t = +\infty$, i.e.\ denotes an integration over a function space (all functions $x(t)$). Since a function has infinitely many degrees of freedom, a path integral can be considered as an integral over an infinite number of variables (labeled by $t$). The observable $x^2$ has to be evaluated on all these paths (e.g.\ at time $t=0$) and ``weighted'' with the phase $e^{i S[x]} / Z$.

The transition from quantum mechanics to quantum field theory, e.g.\ to QCD, is straightforward (hats $\hat{\phantom{x}}$ on top of operators are in the following omitted):
\begin{eqnarray}
\nonumber & & \hspace{-0.7cm} \langle \Omega | \mathcal{O}[\psi,\bar{\psi},A] | \Omega \rangle \ \ = \ \ \frac{1}{Z} \int D\psi \, D\bar{\psi} \, \int DA \, \mathcal{O}[\psi,\bar{\psi},A] e^{i S_\textrm{QCD}[\psi,\bar{\psi},A]} \quad , \\
\label{EQN451} & & \hspace{0.675cm} \quad Z \ \ = \ \ \int D\psi \, D\bar{\psi} \, \int DA \, e^{i S_\textrm{QCD}[\psi,\bar{\psi},A]} ,
\end{eqnarray}
where
\begin{itemize}
\item $| \Omega \rangle$ is the QCD ground state, i.e.\ the vacuum,

\item $\int D\psi \, D\bar{\psi} = \prod_{x_\mu} d\psi(x) \, d\bar{\psi}(x)$ is the integration over all possible quark field configurations $\psi^{(f)}(\mathbf{r},t)$,

\item $\int DA = \prod_{x_\mu} \prod_{\nu = 0}^3 dA_\nu(x)$ is the integration over all possible gluon field configurations $A_\mu(\mathbf{r},t)$,

\item $\mathcal{O}[\psi,\bar{\psi},A]$ is an observable composed of quark and gluon fields/field operators.
\end{itemize}

To determine a hadron mass, one needs to calculate a temporal correlation function of a suitable hadron creation operator. E.g.\ to determine the pion mass,
\begin{eqnarray}
\nonumber & & \hspace{-0.7cm} \langle \Omega | \Big(\mathcal{O}_\pi(t_2)\Big)^\dagger \mathcal{O}_\pi(t_1) | \Omega \rangle \ \ = \ \ \frac{1}{Z} \int D\psi \, D\bar{\psi} \, \int DA \, \Big(\mathcal{O}_\pi(t_2)\Big)^\dagger \mathcal{O}_\pi(t_1) e^{i S_\textrm{QCD}[\psi,\bar{\psi},A]}, \\
 & & \\
 & & \hspace{-0.7cm} \mathcal{O}_\pi(t) \ \ = \ \ \int d^3r \, \bar{u}(\mathbf{r},t) \gamma_5 d(\mathbf{r},t)
\end{eqnarray}
is needed (cf.\ section~\ref{SEC003} for details).

Solving QCD path integrals (\ref{EQN451}) analytically seems to be impossible. Path integrals are, however, suited for numerical evaluation on HPC systems ($\rightarrow$ lattice field theory, lattice gauge theory, lattice QCD), which is one of the reasons, why they are commonly used in quantum field theory.
\end{itemize}


\subsection{\label{SEC877}Lattice QCD (numerical evaluation of QCD path integrals)}

\begin{figure}[htb]
\begin{center}
\input{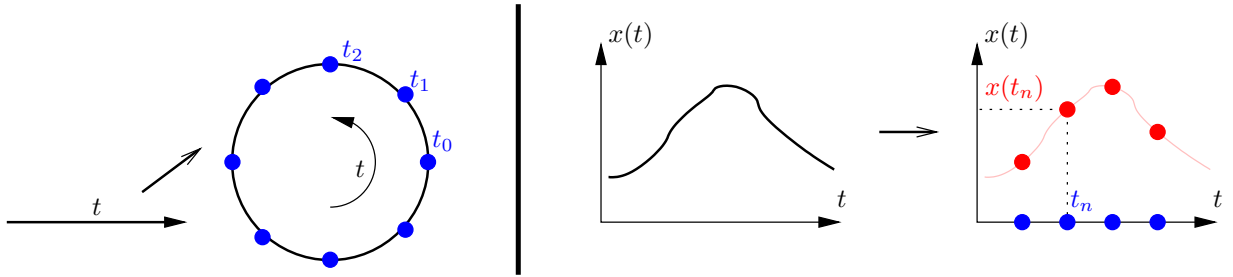}
\caption{\label{FIG005}lattice discretization of a path integral in quantum mechanics.}
\end{center}
\end{figure}

The basic ideas behind lattice computations can most easily be illustrated in quantum mechanics. To be able to solve the infinite dimensional path integral (\ref{EQN891}) numerically, one first has to reduce it to an ordinary multidimensional integral. To this end
\begin{itemize}
\item time is discretized, $t \in \mathbb{R} \quad \rightarrow \quad t_n = n \times a$, $n \in \mathbb{Z}$ ($a$ is the lattice spacing),

\item time is considered to be periodic (periodicity $T = a N_T$, $N_T$ is the number of lattice sites), i.e.\ $t_n \equiv t_{n + N_T}$
\end{itemize}
(cf.\ Figure~\ref{FIG005}, left). The lattice path integral is then
\begin{eqnarray}
\int Dx \ \ \rightarrow \ \ \prod_{n=0}^{N_T-1} \int dx(t_j)
\end{eqnarray}
(cf.\ Figure~\ref{FIG005}, right).

Another severe numerical problem is the oscillating weight $e^{i S[x]}$ in (\ref{EQN891}). To avoid these oscillations, one considers so-called Euclidean path integrals:
\begin{itemize}
\item Time evolution via $e^{-i H t}$ is replaced by $e^{-H t}$ ($t$ is then only a mathematical parameter, which should not be interpreted as physical time).

\item ``Euclidean expressions'' usually differ from the original ``Minkowski expressions'' by signs or phases $\pm i$ and are obtained via $t \rightarrow -i t$ (changing the sign in front of $t^2$ changes Minkowski spacetime [metric $ds^2 = \pm (dt^2 - dx^2)$] to Euclidean spacetime [metric $ds^2 = -(dt^2 + dx^2)$] and vice versa); in particular
\begin{eqnarray}
 & & \hspace{-0.7cm} \int Dx \, e^{i S[x]} \quad , \quad S[x] \ \ = \ \ \int dt \, \bigg(\frac{m}{2} \dot{x}^2 - V(x)\bigg) \quad \textrm{(Minkowski)} \\
\label{EQN774} & & \hspace{-0.7cm} \int Dx \, e^{-S[x]} \quad , \quad S[x] \ \ = \ \ \int dt \, \bigg(\frac{m}{2} \dot{x}^2 + V(x)\bigg) \quad \textrm{(Euclidean)}.
\end{eqnarray}
\end{itemize}
The real positive weight $e^{-S[x]}$ in the Euclidean path integral (\ref{EQN774}) exponentially suppresses paths $x(t)$ with respect to the minimum or the minima of the Euclidean action. This allows for efficient stochastic methods to numerically evaluate such Euclidean path integrals (importance sampling via Monte Carlo algorithms). Note that Minkowski and Euclidean path integrals are not equivalent. Quite often, however, time independent physical observables, in particular energy eigenvalues (in QCD equivalent to hadron masses), can be expressed by Euclidean path integrals.

Performing computations for several small $a$ and several large $T = a N_T$ allows to study and to remove systematic effects due to discretization and periodicity (continuum extrapolation, infinite volume extrapolation).

Lattice computations in QCD follow the same ideas:
\begin{itemize}
\item Spacetime is discretized, $x_\mu \in \mathbb{R}^4 \quad \rightarrow \quad x_\mu = n_\mu \times a$, $n_\mu \in \mathbb{Z}^4$ ($a$ is the lattice spacing); cf.\ Figure~\ref{FIG003}.

\begin{figure}[htb]
\begin{center}
\input{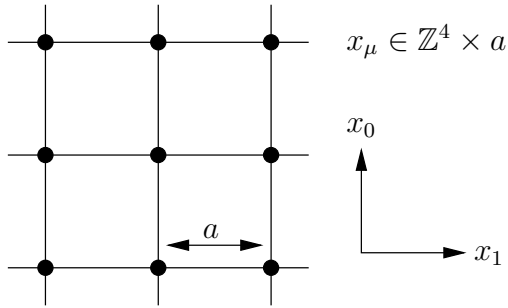}
\caption{\label{FIG003}lattice discretization of spacetime.}
\end{center}
\end{figure}

\item Spacetime is considered to be periodic (periodicity $L = a N_L$, $N_L$ is the number of lattice sites in each spacetime direction, $N_L^4$ lattice sites in total), i.e.\ $x_\mu \equiv x_\mu + L e_\mu^{(\nu)}$ ($e^{(\nu)}$ denotes the unit vector in $\nu$ direction). In other words spacetime has the shape of a four-dimensional torus.

\item The resulting finite dimensional lattice QCD path integral
\begin{eqnarray}
\int D\psi \, D\bar{\psi} \, \int DA \ \ \rightarrow \ \ \prod_{n_\mu} \int d\psi(a n_\mu) \, d\bar{\psi}(a n_\mu) \, \int dU(a n_\mu)
\end{eqnarray}
can be solved numerically, when formulated in Euclidean spacetime.
\end{itemize}

The typical present-day dimensionality of a lattice QCD path integral can easily be estimated:
\begin{itemize}
\item $n_\mu \in \{ 0,1, \ldots ,N_L-1\}^4$: e.g.\ for $N_L = 32$, $32^4 \approx 10^6$ lattice sites.

\item $\psi = \psi_A^{a,(f)}$: $24$ quark degrees of freedom for every flavor (real and imaginary part of $\psi$, color $a = 1, \ldots , 3$, spin $A = 1, \ldots ,4$); $\geq 2$ flavors, i.e.\ at least a $u$ and a $d$ quark field are considered.

\item $U = U_\mu^{a b}$ (the lattice equivalent of $A_\mu^a$): $32$ gluon degrees of freedom (color $a = 1, \ldots , 8$, spin $\mu = 0, \ldots ,3$).

\item In total a $32^4 \times (2 \times 24 + 32) \approx 83 \times 10^6$ dimensional integral.
\end{itemize}
Clearly, standard approaches for numerical integration (like uniform sampling) are not applicable. It is mandatory to use sophisticated algorithms (stochastic integration techniques, so-called Monte-Carlo algorithms) and HPC systems.


\subsection{\label{SEC522}A brief overview of some technical aspects of lattice QCD}

In this section some technical aspects of lattice QCD are outlined. Detailed presentations can be found e.g.\ in the textbooks \cite{Rothe,DeGrand,Gattringer}.

\textbf{Lattice derivatives} \\
Derivatives of a continuous function $f$ have to be replaced by finite differences on a lattice. The discretization of a derivative is not unique. For example one could use
\begin{eqnarray}
\partial_\mu f(x) \ \ \rightarrow \ \ \frac{f(x + a e^{(\mu)}) - f(x)}{a}
\end{eqnarray}
as well as
\begin{eqnarray}
\partial_\mu f(x) \ \ \rightarrow \ \ \frac{f(x + a e^{(\mu)}) - f(x - a e^{(\mu)})}{2 a} .
\end{eqnarray}
Both discretizations fulfill the requirement that in the limit $a \rightarrow 0$ the continuum derivative $\partial_\mu f(x)$ is recovered.

\textbf{The continuum gluon field }$A_\mu(x)$\textbf{ versus lattice link variables }$U_\mu(x)$ \\
A fundamental symmetry of QCD, which is not discussed in these lecture notes, but which is essential for the physics of the strong interactions, is gauge symmetry (cf.\ e.g.\ section~2 of \cite{Gattringer}). It is important to preserve gauge symmetry, when discretizing QCD by means of a spacetime lattice. This is not possible, when discretizing the gluon field in a straightforward way, i.e.\ $A_\mu^a(x) \rightarrow A_\mu^a(n_\mu a)$. To preserve gauge symmetry, the discretized gluon field has to be encoded in form of so-called link variables, which connect neighboring lattice sites $x$ and $x + a e^{(\mu)}$:
\begin{eqnarray}
U_\mu(x) \ \ \approx \ \ \exp\Big(-i g A_\mu(x + a e^{(\mu)}/2) a\Big)
\end{eqnarray}
(note that both $A_\mu$ and $U_\mu$ are $3 \times 3$ matrices in color space). The covariant derivative $D_\mu = \partial_\mu - i g A_\mu$ in the continuum quark action (\ref{EQN755}) depends on the gluon field. Hence, a lattice equivalent contains link variables, e.g.\
\begin{eqnarray}
\label{EQN359} D_\mu f(x) \ \ \to \ \ \frac{U_\mu(x) f(x + a e^{(\mu)}) - U^\dagger(x - a e^{(\mu)}) f(x - a e^{(\mu)})}{2 a}
\end{eqnarray}

\textbf{The lattice quark action} \\
A straightforward discretization of the Euclidean quark action is
\begin{eqnarray}
\nonumber & & \hspace{-0.7cm} S_\textrm{quark}[\psi,\bar{\psi},A] \ \ = \ \ \int d^4x \, \sum_f \bar{\psi}^{(f)} \Big(\gamma_\mu D_\mu + m^{(f)}\Big) \psi^{(f)} \ \ \rightarrow \\
\nonumber & & \rightarrow \ \ \sum_{x_\mu} a^4 \sum_f \bar{\psi}^{(f)}(x) \\
\nonumber & & \hspace{0.675cm} \bigg(\gamma_\mu \frac{U_\mu(x) \psi^{(f)}(x + a e^{(\mu)}) - U_\mu^\dagger(x - a e^{(\mu)}) \psi^{(f)}(x - a e^{(\mu)})}{2 a} + m^{(f)} \psi^{(f)}(x)\bigg) , \\
 & &
\end{eqnarray}
where the spacetime integral has been replaced by a sum over all lattice sites, $\int d^4x \rightarrow \sum_{x_\mu} a^4$. This lattice action, however, leads to an unwanted multiplication of quark flavors (the famous ``fermion doubling problem''), an effect caused by the symmetric discretization of the covariant derivative (\ref{EQN359}). This fermion doubling does not even vanish in the limit $a \rightarrow 0$. There are several possibilities to circumvent the problem, e.g.\ adding the so-called Wilson term $- (a/2) \int d^4x \, \sum_f \bar{\psi}^{(f)} D_\mu D_\mu \psi^{(f)}$ to the quark action (again cf.\ e.g.\ \cite{Rothe,DeGrand,Gattringer} for details).

\textbf{The lattice gluon action} \\
For the lattice gluon action one needs a discretization of the field strength in terms of link variables, which is given by a loop of four link variables:
\begin{eqnarray}
\nonumber & & \hspace{-0.7cm} U_{\mu \nu}(x) \ \ = \ \ U_\mu(x) U_\nu(x + a e^{(\mu)}) U_\mu^\dagger(x + a e^{(\nu)}) U_\nu^\dagger(x) \ \ \approx \\
 & & \approx \ \ \exp\Big(-i g F_{\mu \nu}(x + a e^{(\mu)}/2 + a e^{(\nu)}/2) a^2\Big) .
\end{eqnarray}
A possible discretization of the Euclidean gluon action is
\begin{eqnarray}
S_\textrm{gluon}[A] \ \ = \ \ \frac{1}{2} \int d^4x \, \textrm{Tr}\Big(F_{\mu \nu} F_{\mu \nu}\Big) \ \ \rightarrow \ \ \frac{1}{g^2} \sum_{x_\mu} \sum_{\mu,\nu} \textrm{Tr}\bigg(1 - \frac{1}{2} \Big(U_{\mu \nu}(x) + U_{\mu \nu}^\dagger(x)\Big)\bigg) .
\end{eqnarray}

\textbf{Choosing and setting the lattice spacing }$a$ \\
In lattice QCD computations all dimensionful quantities are expressed in units of the lattice spacing $a$, i.e.\ as dimensionless quantities. For example $\psi^{(f)} \rightarrow \hat{\psi}^{(f)} = \psi^{(f)} a^{3/2}$ and $m^{(f)} \rightarrow \hat{m}^{(f)} = a m^{(f)}$. It is easy to show that after these replacements the lattice spacing $a$ does not appear anymore explicitly in the lattice QCD action. On the other hand, the lattice spacing $a$ and the QCD coupling constant $g$ are related. The functional dependence $g = g(a)$ is non-trivial, but can be determined numerically. Therefore, setting $a$ to a specific desired value amounts to choosing the corresponding value for $g$ (in the literature typically the equivalent parameter $\beta = 6 / g^2$ is used).

The lattice spacing should be chosen in line with the following criteria:
\begin{itemize}
\item The number of lattice sites $N_L^4$, which can be handled in a computation, is limited by the available computer resources. With up-to-date HPC systems typically $24 \ltapprox N_L \ltapprox 64$ is feasible. The physical extension of the lattice in each of the four spacetime directions is $L = a N_L$. $L$ should be chosen large enough, such that the hadron of interest fits into the spatial volume of the lattice. For example, if one studies a hadron with a diameter of around $1 \, \textrm{fm}$, a reasonable choice for the spatial extension of the lattice could be $L \gtapprox 2 \, \textrm{fm}$, which implies $a \gtapprox 2 \, \textrm{fm} / N_L$. Moreover, the pion is the lightest particle in QCD and unwanted effects from periodicity are in most cases dominated by pion exchange. One can show that these effects are exponentially suppressed proportional to $m_\pi L$. A common rule of thumb is to choose $m_\pi L \gtapprox 3 \ldots 5$, which amounts to $a \gtapprox (3 \ldots 5) / m_\pi N_L$.

\item On the other hand, studying the above mentioned hadron with a diameter of around $1 \, \textrm{fm}$ will only be successful on a lattice, which is able to resolve its substructure, i.e.\ if $a \ll 1 \, \textrm{fm}$. Typical nowadays lattice spacings, which lead to rather precise results for many hadron masses, in particular those composed of $u$, $d$ and $s$ quarks, are in the range $0.05\, \textrm{fm} \ltapprox a \ltapprox \textrm{0.15} \, \textrm{fm}$.
\end{itemize}


\section{\label{SEC003}Symmetries of QCD, hadron creation operators, \\ temporal correlation functions}


\subsection{Classification of hadrons}

Hadrons and their properties are compiled by the Particle Data Group (cf.\ \\ \texttt{http://pdg.lbl.gov/}) \cite{PDG}. Hadronic states are mainly classified by QCD quantum numbers:
\begin{itemize}
\item Spin or total angular momentum $J$ (bosons: $J = 0, 1, 2, \ldots$; fermions \\ $J = 1/2, 3/2, 5/2, \ldots$).

\item Parity (spatial reflections) $P = \pm 1$.

\item Charge conjugation (exchange of quarks and antiquarks) $C = \pm 1$ (flavorless mesons only).

\item Flavor quantum numbers:
\\ $\phantom{XXX}$ Isospin: $I$; $I_z = +1/2$ ($u$), $I_z = -1/2$ ($d$).
\\ $\phantom{XXX}$ Strangeness: $S = -1$ ($s$), $S = +1$ ($\bar{s}$).
\\ $\phantom{XXX}$ Charm: $C = +1$ ($c$), $C = -1$ ($\bar{c}$).
\\ $\phantom{XXX}$ Bottomness: $B' = -1$ ($b$), $B' = +1$ ($\bar{b}$).
\\ $\phantom{XXX}$ Topness: $T = +1$ ($t$), $T = -1$ ($\bar{t}$).

\item Since electromagnetism is neglected throughout these lecture notes, electrical charge is not discussed or used in the following.
\end{itemize}

As in quantum mechanics, quantum numbers correspond to eigenvalues of operators, which commute with the Hamiltonian. In other words these operators generate symmetry transformations. Familiar examples are $[H,J^2] = 0$ (QCD is symmetric under rotations) or $[H,P] = 0$ (QCD is symmetric under parity).

A hadron with quantum numbers $I(J^P)$ (and $S$, $C$, $B'$, $T$, which are often not listed) corresponds to a low lying eigenstate of the QCD Hamiltonian with these quantum numbers and its mass to the corresponding eigenvalue\footnote{This statement is only correct for hadrons, which are (essentially) stable with respect to hadronic decays, e.g.\ $\pi$, $K$, $D$, ..., $p$, $n$, ... There are other hadrons, which can (readily) decay into multi-particle states, e.g.\ $\kappa \equiv K_0^\ast(800) \rightarrow K + \pi$ (because of $m_\kappa \approx 672 \, \textrm{MeV} > m_K + m_\pi \approx (495 + 140) \, \textrm{MeV}$, and because both $\kappa$ and $K + \pi$ can have quantum numbers $I(J^P) = 1/2(0^+)$). For such hadrons (so-called resonances) a mass determination is significantly more difficult (cf.\ e.g.\ \cite{Prelovsek:2010kg,Alexandrou:2012rm,Lang:2012sv}).}. For example for the pion, which is characterized by $I(J^P) = 1(0^-)$,
\begin{itemize}
\item $\hat{I}^2 | \pi \rangle = I (I + 1) | \pi \rangle = 2 | \pi \rangle$,

\item $\hat{J}^2 | \pi \rangle = J (J + 1) | \pi \rangle = 0 | \pi \rangle$,

\item $\hat{P} | \pi \rangle = P | \pi \rangle = -| \pi \rangle$,
\end{itemize}
while its mass is given by
\begin{itemize}
\item $\hat{H} | \pi \rangle = E | \pi \rangle = (m_\pi + E_\Omega) | \pi \rangle \quad \rightarrow \quad m_\pi = E - E_\Omega$
\end{itemize}
(this time hats $\hat{\phantom{x}}$ on top of operators are explicitly written, to distinguish operators and eigenvalues/quantum numbers; QCD energy eigenvalues always include $E_\Omega$, the energy of the QCD vacuum, which has quantum numbers $I(J^P) = 0(0^+)$).

Note that quantum numbers do not uniquely characterize a hadron. As an example there are excited versions of the pion: $\pi^0$ ($m_\pi = 135 \, \textrm{MeV}$), $\pi(1300)$ ($m_{\pi(1300)} \approx 1300 \, \textrm{MeV}$) and $\pi(1800)$ ($m_{\pi(1800)} \approx 1812 \, \textrm{MeV}$) share $I(J^P) = 1(0^-)$.

When studying hadrons by means of lattice QCD, typical goals include the following: compute for a given a set of quantum numbers $I(J^P)$ (and $S$, $C$, $B'$, $T$), i.e.\ for a specific hadron (typically the lightest in the given $I(J^P)$ sector)
\begin{itemize}
\item its mass (explained in the following),

\item its structure (not discussed in these lecture notes),

\item decay properties and probabilities (not discussed in these lecture notes).
\end{itemize}


\subsection{Computation of hadron masses}

To determine a hadron mass $m_H$ (the hadron $H$ is specified by quantum numbers $I(J^P)$, ...), one proceeds in two steps:
\begin{itemize}
\item[(1)] Define a suitable hadron creation operator $\mathcal{O}_H$ (cf.\ section~\ref{SEC608}).

\item[(2)] Compute the Euclidean temporal correlation function of the hadron creation operator $\mathcal{O}_H$, then read off the hadron mass $m_H$ from the asymptotic exponential behavior (cf.\ section~\ref{SEC609}).
\end{itemize}


\subsubsection{\label{SEC608}Hadron creation operators}

A hadron creation operator $\mathcal{O}_H$ (also called interpolating operator) is an operator, which, when acting on the QCD vacuum $| \Omega \rangle$, creates a so-called trial state $| \phi \rangle = \mathcal{O}_H | \Omega \rangle$ with the quantum numbers of the hadron $H$, i.e.\ a state with $I(J^P)$, ...

In general $| \phi \rangle$ is not the hadron of interest, i.e.\ $| \phi \rangle \neq | H \rangle$. It is a linear superposition of all states (hadron and multi-hadron states) with quantum numbers $I(J^P)$, ...,
\begin{eqnarray}
\label{EQN234} | \phi \rangle \ \ = \ \ \mathcal{O}_H | \Omega \rangle \ \ = \ \ \sum_{n=0}^\infty a_n | I(J^P) ; n \rangle
\end{eqnarray}
(in the following labels of states are abbreviated according to $| n \rangle \equiv | I(J^P) ; n \rangle$; moreover, the states are ordered according to their masses, i.e.\ $m_0 \leq m_1 \leq m_2 \leq \ldots$). The coefficient $a_n = \langle n | \mathcal{O}_H | \Omega \rangle$ is the overlap of the trial state and the energy eigenstate $| n \rangle$. Its magnitude indicates, to what extent the hadron creation operator $\mathcal{O}_H$ excites the hadronic state $| n \rangle$. Quite often one is interested in the lightest hadronic state in a given sector, i.e.\ $| H \rangle = | 0 \rangle$.

\subsubsection*{Example: pion, quantum numbers $I(J^P) = 1(0^-)$}

The pion is the ground state in the $I(J^P) = 1(0^-)$ sector:
\begin{eqnarray}
| H \rangle \ \ = \ \ | \pi \rangle \ \ = \ \ | 0 \rangle .
\end{eqnarray}
In a simplified version of QCD, where quark antiquark pair creation and annihilation is forbidden\footnote{Calculations in such a simplified world (in the so-called quenched approximation) are computationally much cheaper and are quite common in the older literature.}, the excited states are the previously mentioned excited versions of the pion,
\begin{eqnarray}
\label{EQN235} | n \rangle \ \ \in \ \ \Big\{ | \pi(1300) \rangle \ , \ | \pi(1800) \rangle \ , \ \ldots \Big\} \quad , \quad n \geq 1 .
\end{eqnarray}
In full QCD, where quark antiquark pair creation and annihilation takes place, the low lying energy eigenstates usually contain multi-particle states. For example in the pion sector the first excited state is a three-pion state,
\begin{eqnarray}
\label{EQN236} | n \rangle \ \ \in \ \ \Big\{ | \pi + \pi + \pi \rangle \ , \ \ldots \Big\} \quad , \quad n \geq 1 .
\end{eqnarray}

Hadron creation operators are far from unique. From a numerical point of view a good operator to excite the ground state corresponds to $|a_n| / |a_0| \approx 0$, $n \geq 1$. In other words $\mathcal{O}_H$ excites the quark fields and the gluon field in such a way that $\mathcal{O}_H | \Omega \rangle$ closely resembles $| H \rangle = | 0 \rangle$.

A typical hadron creation operator for the pion is
\begin{eqnarray}
\mathcal{O}_\pi \ \ = \ \ \int d^3r \, \bar{u}(\mathbf{r}) \gamma_5 d(\mathbf{r}) .
\end{eqnarray}
\begin{itemize}
\item $\bar{u}(\mathbf{r}) d(\mathbf{r})$ realizes $I = 1$.

\item $\gamma_5$ realizes $J^P = 0^-$.

\item $\int d^3r$ realizes momentum $\mathbf{p} = 0$ (without $\int d^3r$ hadrons with non-vanishing momenta would appear in (\ref{EQN234}), (\ref{EQN235}) and (\ref{EQN236}).
\end{itemize}
For detailed explanations and further examples cf.\ section~\ref{SEC458}.


\subsubsection{\label{SEC609}Temporal correlation functions}

The vacuum expectation value of a hadron creation operator $\mathcal{O}_H$ at time $t_1$ and its hermitian conjugate $\mathcal{O}_H^\dagger$ at time $t_2$ is called a temporal correlation function. Temporal correlation functions are common representatives of quantities, which are computed in lattice QCD. They are defined via
\begin{eqnarray}
\nonumber & & \hspace{-0.7cm} C_H(\Delta t) \ \ = \ \ \langle \Omega | \mathcal{O}_H^\dagger(t_2) \mathcal{O}_H(t_1) | \Omega \rangle \ \ = \ \ \frac{1}{Z} \int D\psi \, D\bar{\psi} \, \int DA \, \mathcal{O}_H^\dagger(t_2) \mathcal{O}_H(t_1) e^{-S_\textrm{QCD}[\psi,\bar{\psi},A]} , \\
 & &
\end{eqnarray}
where $\Delta t = t_2 - t_1$ \footnote{For technical aspects, i.e.\ detailed discussions, how a such temporal correlation functions can be computed using lattice QCD, cf.\ e.g.\ \cite{Rothe,DeGrand,Gattringer}.}.

It is easy to show that a temporal correlation function is dominated by the ground state for large $\Delta t$. The reason is that states are exponentially suppressed proportional to their mass and the temporal separation $\Delta t$:
\begin{eqnarray}
\nonumber & & \hspace{-0.7cm} C_H(\Delta t) \ \ = \ \ \langle \Omega | \mathcal{O}_H^\dagger(t_2) 
\mathcal{O}_H(t_1) | \Omega \rangle \ \ = \ \ \sum_{n=0}^\infty \langle \Omega | 
\mathcal{O}_H^\dagger(t_2) | n \rangle \langle n | \mathcal{O}_H(t_1) | \Omega \rangle \ \ = \\
\nonumber & & = \ \ \sum_{n=0}^\infty \langle \Omega | e^{+H \Delta t} \mathcal{O}_H^\dagger(t_1) 
e^{-H \Delta t} | n \rangle \langle n | \mathcal{O}_H(t_1) | \Omega \rangle \ \ = \\
\label{EQN650} & & = \ \ \sum_{n=0}^\infty \underbrace{\Big|\langle n | \mathcal{O}_H | \Omega \rangle
 \Big|^2}_{= |a_n|^2} \exp\Big(-\underbrace{(E_n - E_\Omega)}_{= m_n} \Delta t\Big) \ \ 
 \stackrel{\Delta t \rightarrow \infty}{=} \ \ |a_0|^2 e^{-m_0 \Delta t} ,
\end{eqnarray}
where the Euclidean time evolution explained in section~\ref{SEC877} and (\ref{EQN234}) has been used.

\begin{figure}[htb]
\begin{center}
\input{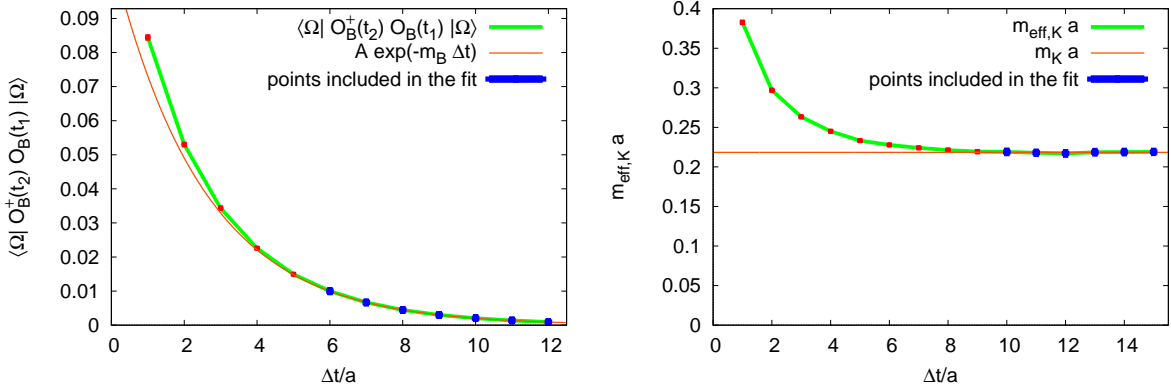}
\caption{\label{FIG001}
\textbf{left}: the temporal correlation function of a $B$ meson creation operator as a function of the temporal separation (taken from \cite{Blossier:2009vy});
\textbf{right}: the effective mass of the kaon as a function of the temporal separation (taken from \cite{Baron:2010th}).
}
\end{center}
\end{figure}

To determine the ground state hadron mass $m_H = m_0$, one simply has to fit a function $A e^{-m_H \Delta t}$ (fitting parameters $A$ and $m_H$) to the lattice results for the temporal correlation function $C_H(\Delta t)$ at sufficiently large $\Delta t$ (an example is shown in Figure~\ref{FIG001}, left).

In practice the hadron mass $m_H$ is often determined from a related quantity, the so-called effective mass, which is given by
\begin{eqnarray}
m_{\textrm{eff},H}(\Delta t) \ \ = \ \ \frac{1}{a} \log\bigg(\frac{C_H(\Delta t)}{C_H(\Delta t+a)}\bigg) .
\end{eqnarray}
Inserting (\ref{EQN650}) leads to
\begin{eqnarray}
\nonumber & & \hspace{-0.7cm} m_{\textrm{eff},H}(\Delta t) \ \ = \ \ \frac{1}{a} \log\bigg(\frac{\sum_{n=0}^\infty |a_n|^2 e^{-m_n \Delta t}}{\sum_{n=0}^\infty |a_n|^2 e^{-m_n (\Delta t+a)}}\bigg) \ \ = \\
\label{EQN850} & & = \ \ \frac{1}{a} \log\bigg(e^{+m_H a} \underbrace{\frac{1 + \sum_{n=1}^\infty \frac{|a_n|^2}{|a_0|^2} e^{-(m_n - m_H) \Delta t}}{1 + \sum_{n=1}^\infty \frac{|a_n|^2}{|a_0|^2} e^{-(m_n - m_H) (\Delta t + a)}}}_{= 1 + \mathcal{O}(e^{-(m_1 - m_H) \Delta t})}\bigg) \ \ \stackrel{\Delta t \rightarrow \infty}{=} \ \ m_H .
\end{eqnarray}
The effective mass becomes a constant in the limit $\Delta t \rightarrow \infty$, which is the ground state hadron mass $m_H = m_0$. To determine $m_H$, one simply has to fit a constant to the lattice results for the effective mass $m_{\textrm{eff},H}(\Delta t)$ at sufficiently large $\Delta t$ (an example is shown in Figure~\ref{FIG001}, right).

Note that (\ref{EQN650}) to (\ref{EQN850}) are only correct, if the temporal direction is infinitely extended. For a finite periodic spacetime lattice (periodicity $L$), these expressions are more complicated. For example the analog of (\ref{EQN650}) is $C_H(\Delta t) \stackrel{\Delta t \approx L/2}{\approx} |a_0|^2 (e^{-m_0 \Delta t} + e^{-m_0 (L - \Delta t)})$. For more details cf.\ e.g.\ \cite{Creutz:1976ch}.


\section{\label{SEC458}More about hadron creation operators, exemplary \\ lattice results}

The intention of this section is to repeat, to illustrate and to extend the basics of lattice hadron spectroscopy in the context of recent lattice papers.


\subsection{\label{SEC894}Example~1: the spectrum of $D$ and $D_s$ mesons and of \\ charmonium \cite{Kalinowski:2012re,Kalinowski:2013wsa}}

The goals of this example are to present elements of an ongoing straightforward lattice QCD meson spectroscopy project and to discuss quantum numbers of mesonic trial states in more detail.

\begin{figure}[htb]
\begin{center}
\input{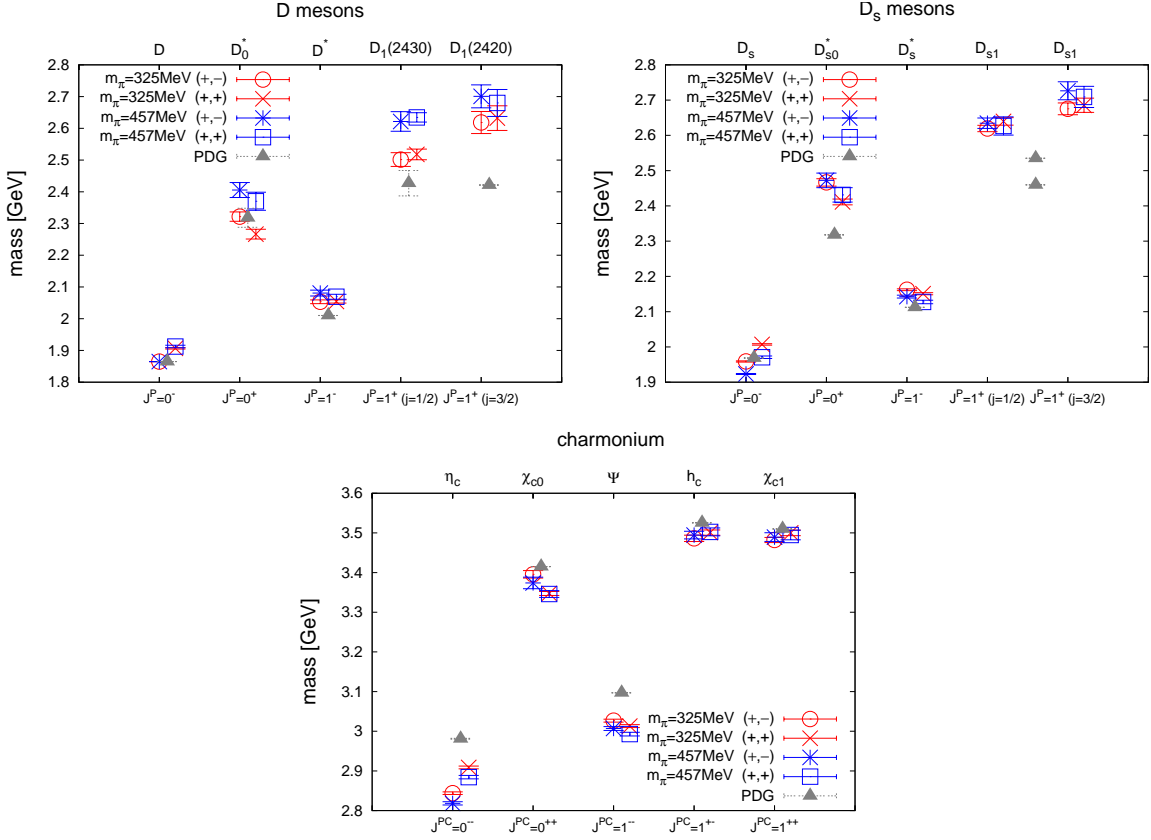}
\caption{\label{FIG002}the spectrum of $D$ and $D_s$ mesons and of charmonium (taken from \cite{Kalinowski:2012re,Kalinowski:2013wsa}); red data points: lattice results for light $u/d$ quark masses corresponding to $m_\pi \approx 325 \, \textrm{MeV}$ (circles and crosses distinguish two different lattice discretizations); blue data points: lattice results for light $u/d$ quark masses corresponding to $m_\pi \approx 457 \, \textrm{MeV}$ (stars and boxes distinguish two different lattice discretizations); gray data points: experimental results \cite{PDG}.}
\end{center}
\end{figure}

Several $D \equiv \bar{c} u$, $D_s \equiv \bar{c} s$ and $\textrm{charmonium} \equiv \bar{c} c$ states have experimentally been observed \cite{PDG}. Some of them are plotted in Figure~\ref{FIG002} together with corresponding lattice results \cite{Kalinowski:2012re,Kalinowski:2013wsa}. Suitable $D$ meson creation operators, which have been used, to obtain these lattice results, are
\begin{eqnarray}
\label{EQN378} D \ \ (J^P = 0^-) \quad & \rightarrow & \quad \mathcal{O}_D \ \ = \ \ \int d^3r \, \bar{c}(\mathbf{r}) \gamma_5 u(\mathbf{r}) , \\
\label{EQN380} D_0^\ast \ \ (J^P = 0^+) \quad & \rightarrow & \quad \mathcal{O}_{D_0^\ast} \ \ = \ \ \int d^3r \, \bar{c}(\mathbf{r}) u(\mathbf{r}) , \\
\label{EQN381} D^\ast \ \ (J^P = 1^-) \quad & \rightarrow & \quad \mathcal{O}_{D^\ast} \ \ = \ \ \int d^3r \, \bar{c}(\mathbf{r}) \gamma_j u(\mathbf{r}) \quad , \quad j = 1,2,3 , \\
\label{EQN379} D_1 \ \ (J^P = 1^+) \quad & \rightarrow & \quad \mathcal{O}_{D_1} \ \ = \ \ \int d^3r \, \bar{c}(\mathbf{r}) \gamma_j \gamma_5 u(\mathbf{r}) \quad , \quad j = 1,2,3 .
\end{eqnarray}
In the following the quantum numbers of the corresponding trial states $\mathcal{O}_H | \Omega \rangle$ are explained in detail.

\subsubsection*{Flavor quantum numbers}

\begin{itemize}
\item $I = 1/2$, $C = \pm 1$, i.e.\ flavor is trivial. One needs a $c$ antiquark and a light $u$ or $d$ quark or vice versa, i.e.\ $\bar{c} u$, $\bar{c} d$, $\bar{u} c$ or $\bar{d} c$.
\end{itemize}

\subsubsection*{Parity}

\begin{itemize}
\item Using relativistic quantum mechanics or field theory one can show \\
$\psi(\mathbf{r}) \rightarrow_P P(\psi(\mathbf{r})) = \gamma_0 \psi(-\mathbf{r})$.

\item Consequently, \\
$\bar{\psi}(\mathbf{r}) = \psi^\dagger(\mathbf{r}) \gamma_0 \rightarrow_P P(\psi^\dagger(\mathbf{r}) \gamma_0) = \psi^\dagger(-\mathbf{r}) \gamma_0 \gamma_0 = \bar{\psi}(-\mathbf{r}) \gamma_0$.

\item The QCD vacuum has positive parity, i.e.\ \\
$| \Omega \rangle \rightarrow_P P(| \Omega \rangle) = | \Omega \rangle$.

\item Using these relations one can show that the $D$ meson creation operators (\ref{EQN378}) to (\ref{EQN379}) create trial states with the listed parity, e.g.\
\begin{eqnarray}
\nonumber & & \hspace{-0.7cm} \int d^3r \, \bar{c}(\mathbf{r}) \gamma_5 u(\mathbf{r}) | \Omega \rangle \ \ \rightarrow_P \ \ P\bigg(\int d^3r \, \bar{c}(\mathbf{r}) \gamma_5 u(\mathbf{r}) | \Omega \rangle\bigg) \ \ = \\
 & & = \ \ \underbrace{\int d^3r}_{= \int d^3r^\prime} \, \bar{c}(\underbrace{-\mathbf{r}}_{= \mathbf{r^\prime}}) \underbrace{\gamma_0 \gamma_5 \gamma_0}_{= -\gamma_5} u(\underbrace{-\mathbf{r}}_{= \mathbf{r^\prime}}) | \Omega \rangle \ \ = \ \ -\int d^3r \, \bar{c}(\mathbf{r}) \gamma_5 u(\mathbf{r}) | \Omega \rangle \\
\nonumber & & \hspace{-0.7cm} \int d^3r \, \bar{c}(\mathbf{r}) u(\mathbf{r}) | \Omega \rangle \ \ \rightarrow_P \ \ P\bigg(\int d^3r \, \bar{c}(\mathbf{r}) u(\mathbf{r}) | \Omega \rangle\bigg) \ \ = \\
 & & = \ \ \underbrace{\int d^3r}_{= \int d^3r^\prime} \, \bar{c}(\underbrace{-\mathbf{r}}_{= \mathbf{r^\prime}}) \underbrace{\gamma_0 \gamma_0}_{= 1} u(\underbrace{-\mathbf{r}}_{= \mathbf{r^\prime}}) | \Omega \rangle \ \ = \ \ +\int d^3r \, \bar{c}(\mathbf{r}) u(\mathbf{r}) | \Omega \rangle .
\end{eqnarray}
\end{itemize}

\subsubsection*{Angular momentum}

\begin{itemize}
\item Using relativistic quantum mechanics or field theory one can show
\begin{itemize}
\item $\bar{\psi} \Gamma \psi$, $\Gamma \in \{ 1, \gamma_0 , \gamma_5 , \gamma_0 \gamma_5 \}$ is invariant under rotations,

\item $\bar{\psi} \Gamma_j \psi$, $j=1,2,3$, $\Gamma_j \in \{ \gamma_j , \gamma_0 \gamma_j , \gamma_j \gamma_5 , \gamma_0 \gamma_j \gamma_5 \}$ transform under rotations as the components of an ordinary 3-vector, e.g.\ $(x,y,z)$.
\end{itemize}

\item (A specific variant of the) Wigner-Eckart theorem: if a set of $2 J + 1$ operators $\mathcal{O}_{J M}$, $M = -J,-J+1,\ldots,+J-1,+J$ transforms under rotations as the spherical harmonics $Y_{J M}$ ($\mathcal{O}_{J M}$ are then called spherical tensor operators), the trial states $\mathcal{O}_{J M} | \Omega \rangle$ have total angular momentum $J$ and $z$-component of total angular momentum $M$.

\item Since $Y_{0 0}$ is invariant under rotations, the $D$ meson creation operators $\int d^3r \, \bar{c}(\mathbf{r}) \gamma_5 u(\mathbf{r})$ and $\int d^3r \, \bar{c}(\mathbf{r}) u(\mathbf{r})$ yield spin $J = 0$.

\item Since $Y_{1 M}$, $M = -1,0,+1$ are proportional to $x$, $y$ and $z$ and, therefore, transform under rotations as the components of an an ordinary 3-vector, the $D$ meson creation operators $\int d^3r \, \bar{c}(\mathbf{r}) \gamma_j u(\mathbf{r})$ and $\int d^3r \, \bar{c}(\mathbf{r}) \gamma_j \gamma_5 (\mathbf{r})$ yield spin $J = 1$.
\end{itemize}

\subsubsection*{Momentum}

\begin{itemize}
\item A Fourier transform $\int d^3r \, e^{-i \mathbf{k} \mathbf{r}} \ldots$ of a spatially localized operator $\hat{\mathcal{O}}(\mathbf{r})$ yields momentum $\mathbf{k}$, when applied to the vacuum:
\begin{eqnarray}
\nonumber & & \hspace{-0.7cm} \int d^3r \, e^{-i \mathbf{k} \mathbf{r}} \hat{\mathcal{O}}(\mathbf{r}) | \Omega \rangle \ \ = \\
\nonumber & & = \ \ \int d^3r \, e^{-i \mathbf{k} \mathbf{r}} \int d^3p \sum_n | \mathbf{p} ; n \rangle \langle \mathbf{p} ; n | e^{+i \hat{\mathbf{p}} \mathbf{r}} \hat{\mathcal{O}}(\vec{0}) \underbrace{e^{-i \hat{\mathbf{p}} \mathbf{r}} | \Omega \rangle}_{= | \Omega \rangle} \ \ = \\
\nonumber & & = \ \ \int d^3p \sum_n | \mathbf{p} ; n \rangle \underbrace{\int d^3r \, e^{i (\mathbf{p} - \mathbf{k}) \mathbf{r}}}_{= (2 \pi)^3 \delta(\mathbf{p} - \mathbf{k})} \langle \mathbf{p} ; n | \hat{\mathcal{O}}(\vec{0}) | \Omega \rangle \ \ = \\
 & & = \ \ \sum_n | \mathbf{k} ; n \rangle (2 \pi)^3 \langle \mathbf{k} ; n | \hat{\mathcal{O}}(\vec{0}) | \Omega \rangle
\end{eqnarray}
(this time hats $\hat{\phantom{x}}$ on top of operators are explicitly written, to distinguish momentum operators from corresponding eigenvalues), where $| \mathbf{p} ; n \rangle$ denote momentum eigenstates with momentum $p$ and $n$ the remaining quantum numbers and labels, i.e.\ $n \equiv (I,J,P,\ldots)$.

\item Consequently, the above $D$ meson creation operators, which all contain $\int d^3 p$, excite a linear superposition of $\mathbf{p} = 0$ states, which also has $\mathbf{p} = 0$.
\end{itemize}

It is interesting to mention some more details regarding the lattice results for $D$ and $D_s$ mesons and for charmonium \cite{Kalinowski:2012re,Kalinowski:2013wsa} shown in Figure~\ref{FIG002}:
\begin{itemize}
\item As mentioned in section~\ref{SEC522} the lattice discretization of e.g.\ derivatives and, therefore, also of the QCD action is not unique. The results shown in Figure~\ref{FIG002} have been obtained with two variants (denoted by $(+,-)$ and $(+,+)$) of the Wilson twisted mass discretization of QCD \cite{Baron:2010bv}. In the limit $a \rightarrow 0$ both discretizations should yield identical results.

\item To perform the limit $a \rightarrow 0$ (also called the continuum limit), computations for several small values of the lattice spacing $a$ are needed. The results from Figure~\ref{FIG002} have been obtained at a single lattice spacing $a \approx 0.086 \, \textrm{fm}$. Consequently, a continuum extrapolation could not yet be performed.

\item Light quarks are computationally extremely expensive. Hence, it is common to perform computations not at physically light $u$ and $d$ quark masses, but at several unphysically heavy $u$ and $d$ mass values. The results are then used to perform an extrapolation to the so-called physical point, i.e.\ to the physical value of the $u$ and $d$ quark mass. For the results shown in Figure~\ref{FIG002} the $u$ and $d$ quark mass corresponds to $m_\pi \approx 325 \, \textrm{MeV}$ and to $m_\pi \approx 457 \, \textrm{MeV}$ (an extrapolation to the physical point [$m_\pi \approx 135 \, \textrm{MeV}$] has not yet been performed).
\end{itemize}

For more lattice literature on $D$ and $D_s$ mesons and on charmonium cf.\ \cite{Mohler:2011ke,Namekawa:2011wt,Liu:2012ze,Dowdall:2012ab,Mohler:2012na,Bali:2012ua,Moir:2013ub}.


\subsection{Example~2: the spectrum of $B$ mesons \cite{Jansen:2008ht,Jansen:2008si,Michael:2010aa}}

The goals of this example are to discuss creation operators for mesons with total angular momentum $J \geq 2$ and to illustrate, how the above mentioned systematic errors (lattice discretization errors, errors due to unphysical quark masses) can be removed by suitable extrapolations.

\begin{figure}[htb]
\begin{center}
\input{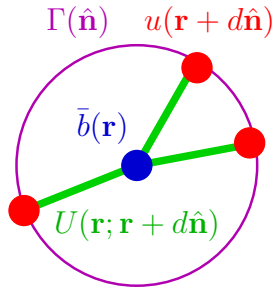}
\caption{\label{FIG004}a $B$ meson creation operator, where the quark and the antiquark can have non-vanishing relative angular momentum.}
\end{center}
\end{figure}

The quark content of $B$ mesons is $\bar{b} u$, $\bar{b} d$, $\bar{u} b$ or $\bar{d} b$. To construct trial states with total angular momentum $J \geq 2$, the two spins of the quark and the antiquark are not sufficient, since they can only be coupled to either $J = 0$ or $J = 1$. Hence, it is necessary to consider a quark-antiquark pair with relative angular momentum, which can be realized by using spherical harmonics. This is, however, only possible, if quark and antiquark are spatially separated. Suitable $B$ meson creation operators are
\begin{eqnarray}
\mathcal{O}_{B\textrm{ meson},\Gamma} \ \ = \ \ \int d^3r \, \bar{b}(\mathbf{r}) \int d\hat{\mathbf{n}} \, \Gamma(\hat{\mathbf{n}}) U(\mathbf{r};\mathbf{r}+d \hat{\mathbf{n}}) u(\mathbf{r}+d \hat{\mathbf{n}})
\end{eqnarray}
(cf.\ Figure~\ref{FIG004}).
\begin{itemize}
\item An $\bar{b}$ quark is created at $\mathbf{r}$, which is the center of the $B$ meson.

\item $\int d\hat{\mathbf{n}}$ denotes the integration over a sphere, which is centered around $\mathbf{r}$, and on which the $u$ quark field is excited.

\item $U(\mathbf{r};\mathbf{r}+d \hat{\mathbf{n}}) = P(\exp(-i g \int_{\mathbf{r}}^{\mathbf{r}+d \hat{\mathbf{n}}} d\mathbf{z} \, \mathbf{A}(\mathbf{z})))$ connects the quark and the antiquark in a gauge invariant way (as already mentioned in section~\ref{SEC522}, gauge symmetry is an essential symmetry of QCD; in particular, hadron creation operators need to be gauge invariant; cf.\ e.g.\ \cite{Rothe,DeGrand,Gattringer} for details). Physically this corresponds to a straight line between $\mathbf{r}$ and $\mathbf{r}+d \hat{\mathbf{n}}$, along which the gluon field is excited. $P$ denotes ``path ordering'' of the $3 \times 3$ color matrices $\mathbf{A}(\mathbf{z})$, i.e., when expanding the exponential, $\mathbf{A}(\mathbf{z}_1)$ appears left of $\mathbf{A}(\mathbf{z}_2)$, if $\mathbf{z}_1$ is closer to $\mathbf{r}$ than $\mathbf{z}_2$.

\item $\Gamma(\hat{\mathbf{n}})$ is a suitable combination of spherical harmonics $Y_{j m}$, \\ $m = -j,-(j-1), \ldots, +(j-1),+j$, and $\gamma$ matrices realizing, according to the Wigner-Eckart theorem, angular momentum $J$ and parity $P$ (cf.\ Table~\ref{Tab001}):
\begin{itemize}
\item The $J = 0$ operators are invariant under rotations, as is the spherical 
harmonic $Y_{0 0}=1/\sqrt{4\pi}$.

\item The $J = 2$ operators transform under rotations as $x^2 - y^2$, which is a specific linear combination of spherical harmonics $Y_{2 M}$.

\item The $J = 3$ operators transform under rotations as $x y z$, which is a specific linear combination of spherical harmonics $Y_{3 M}$.
\end{itemize}

\begin{table}[htb]
\begin{center}

\begin{tabular}{|c|c|c|}
\hline
 & & \vspace{-0.40cm} \\
$\Gamma(\hat{\mathbf{n}})$ &  $J^P$ & notation in Figure~\ref{FIG006} \\
 & & \vspace{-0.40cm} \\
\hline
 & & \vspace{-0.41cm} \\
\hline
 & & \vspace{-0.40cm} \\
$\gamma_5 \ , \ \gamma_5 \gamma_j \hat{n}_j$ & $0^-$ & $S$ \\
 & & \vspace{-0.40cm} \\
$1 \ , \ \gamma_j \hat{n}_j$ &  $0^+$ & $P_-$ \\
 & & \vspace{-0.40cm} \\
\hline
 & & \vspace{-0.40cm} \\
$\gamma_1 \hat{n}_1 - \gamma_2 \hat{n}_2$ (and cyclic) & $2^+$ & $P_+$ \\
 & & \vspace{-0.40cm} \\
$\gamma_5 (\gamma_1 \hat{n}_1 - \gamma_2 \hat{n}_2)$ (and cyclic) & $2^-$ & $D_\pm$ \\
 & & \vspace{-0.40cm} \\
\hline
 & & \vspace{-0.40cm} \\
$\gamma_1 \hat{n}_2 \hat{n}_3 + \gamma_2 \hat{n}_3 \hat{n}_1 + \gamma_3 \hat{n}_1 \hat{n}_2$ & $ 3^-$ & $D_+$ \\
 & & \vspace{-0.40cm} \\
$\gamma_5 (\gamma_1 \hat{n}_2 \hat{n}_3 + \gamma_2 \hat{n}_3 \hat{n}_1 + \gamma_3 \hat{n}_1 \hat{n}_2)$ & $3^+$ & $F_\pm$\vspace{-0.40cm} \\
 & & \\
\hline
\end{tabular}

\caption{\label{Tab001}$B$ meson creation operators and their quantum numbers.}
\end{center}
\end{table}
\end{itemize}

In the six plots in Figure~\ref{FIG006} lattice results for six meson mass differences are shown \cite{Michael:2010aa}: 
\begin{figure}[p]
\begin{center}
\input{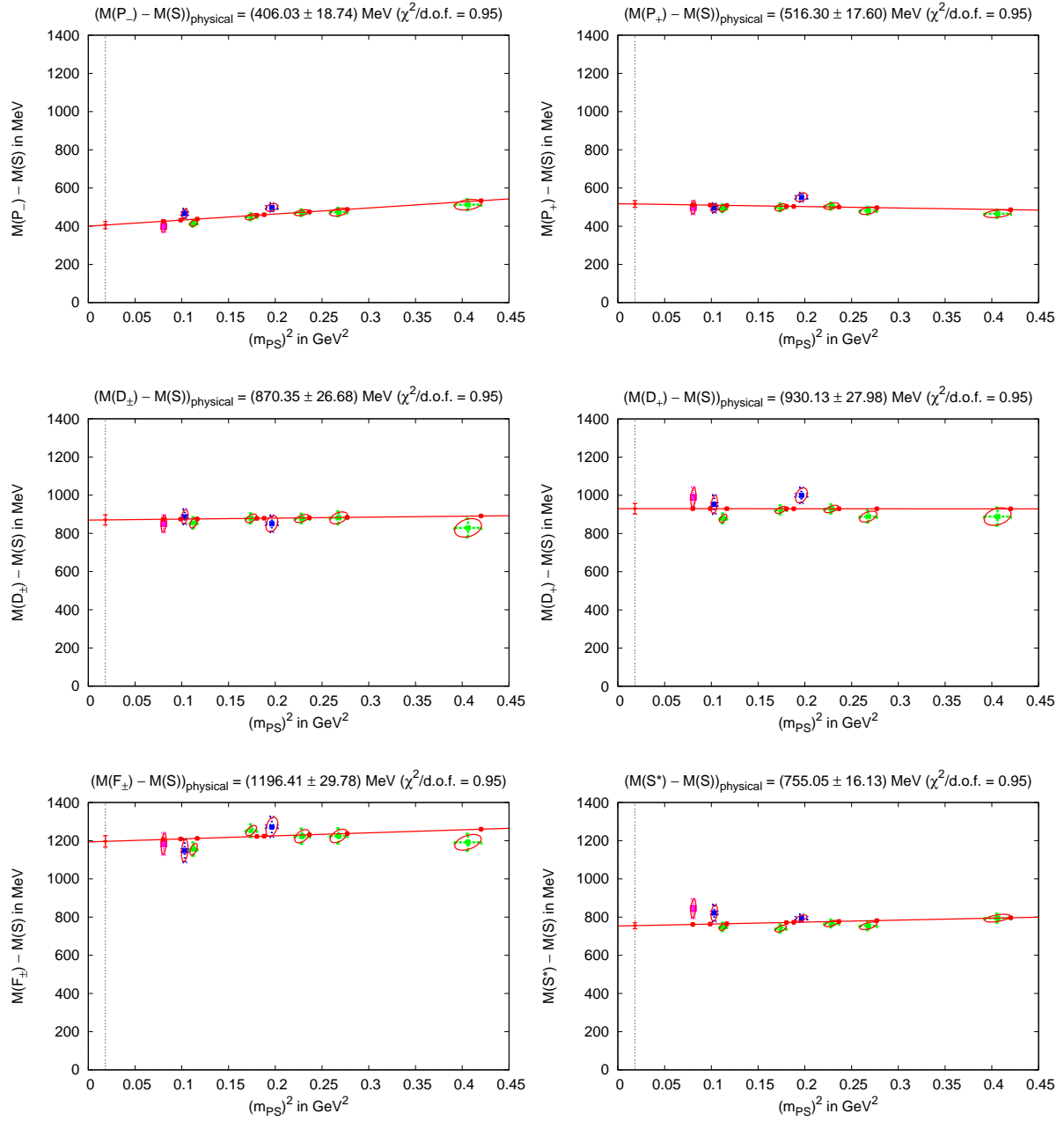}
\caption{\label{FIG006}static-light meson mass differences linearly extrapolated to the physical $u$ quark mass (taken from \cite{Michael:2010aa}).}
\end{center}
\end{figure}
\begin{itemize}
\item The $b$ quark is treated as infinitely heavy, i.e.\ in the static approximation. Since a meson mass is the sum of its constituent quark masses and their binding energy, such static-light mesons, which contain an infinitely heavy quark, are also infinitely heavy, i.e.\ $m_X = E_X - E_\Omega = \infty$ ($X$ labels the meson, e.g.\ $X \in \{ S , P_-, P_+ , \ldots \}$; cf.\ Table~\ref{Tab001}). Therefore, only mass differences between $B$ mesons with distinct internal structure (e.g.\ with different spin $J$ and parity $P$) can be computed, because the infinite quark masses cancel in such meson mass differences. Usually the difference to the lightest static-light meson with quantum numbers $J^P = 0^-$ is considered. In other words, the hierarchy of static-light meson states with different quantum numbers $J^P$ and their level spacings can be computed, despite their infinite mass. Note that static-light mesons are in many respects similar to the hydrogen atom with infinite proton mass, which is well known from standard lectures on quantum mechanics. There, for example, the ground state energy is $E_{0,m_p=\infty} = -13.6 \, \textrm{eV}$ (with the constituent masses, in particular the infinite proton mass, not included), while in a slightly more evolved calculation with finite proton mass one finds $E_{0,\textrm{finite }m_p} = m_p + m_e + E_{0,m_p=\infty} m_p / (m_p+m_e)$.

\item The mass differences are differences to the lightest static-light meson (denoted by $S$), i.e.\ $m_X - m_S$, where $X \in \{ P_-, P_+, D_\pm, D_+, F_\pm, S^\ast \}$ (cf.\ Table~\ref{Tab001}; $S^\ast$ is the first excited state in the $J^P = 0^-$ sector).

\item The horizontal axis corresponds to the light quark mass $m^{(u)}$ (one can show $m^{(u)} \propto m_\pi^2$; notation in Figure~\ref{FIG006}: $m_\textrm{PS} \equiv m_\pi$):
\begin{itemize}
\item Due to limited HPC resources all computations have been performed with unphysically heavy $u$ quarks (cf.\ also section~\ref{SEC894}).

\item The physical point corresponding to $m_\pi^2 = (135 \, \textrm{MeV})^2$ is indicated by the vertical dashed lines in Figure~\ref{FIG006}.

\item Within statistical errors the lattice results behave linearly in $m_\pi^2$. Therefore, the extrapolations to the physical $u$ quark mass have been performed with straight lines (the red lines in Figure~\ref{FIG006}).
\end{itemize}

\item Different colors represent computations at different lattice spacings:
\begin{itemize}
\item Green: $a \approx 0.080 \, \textrm{fm}$.

\item Blue: $a \approx 0.064 \, \textrm{fm}$.

\item Magenta: $a \approx 0.051 \, \textrm{fm}$.

\item Within statistical errors all lattice results are consistent with a single straight line. Lattice discretization errors seem to be negligible, i.e.\ the continuum limit has been reached within statistical errors.
\end{itemize}
\end{itemize}

For more lattice literature on $B$ mesons cf.\ \cite{Michael:1998sg,Green:2003zza,Burch:2006mb,Foley:2007ui,Koponen:2007nr,Burch:2008qx}.


\subsection{Example~3: the spectrum of $b$ baryons \cite{Wagner:2010hj,Wagner:2011fs}}

The goal of this example is to briefly discuss creation operators for baryons.

A $b$ baryon consists of one heavy and two light quarks, i.e.\ $b l l$, where $l \in \{ u , d , s \}$, while an antibaryon requires antiquarks, i.e.\ $\bar{b} \bar{l} \bar{l}$. Suitable $b$ baryon creation operators are
\begin{eqnarray}
\mathcal{O}_{b\textrm{ baryon},\Gamma,\psi^{(1)} \psi^{(2)}} \ \ = \ \ \int d^3r \, \epsilon^{a b c} b^a(\mathbf{r}) \Big((\psi^{b,(1)}(\mathbf{r}))^T \mathcal{C} \Gamma \psi^{c,(2)}(\mathbf{r})\Big) .
\end{eqnarray}
The light quarks form a so-called diquark. The definition of the diquark uses the charge conjugation matrix $\mathcal{C} = \gamma_0 \gamma_2$. Using relativistic quantum mechanics or field theory one can show
\begin{itemize}
\item $\psi^T \mathcal{C} \Gamma \psi$, $\Gamma \in \{ 1, \gamma_0 , \gamma_5 , \gamma_0 \gamma_5 \}$ 
is invariant under rotations,

\item $\psi^T \mathcal{C} \Gamma_j \psi$, $j=1,2,3$, $\Gamma_j \in \{ \gamma_j , \gamma_0 \gamma_j , 
\gamma_j \gamma_5 , \gamma_0 \gamma_j \gamma_5 \}$ transform under rotations as the components 
of an ordinary 3-vector, e.g.\ $(x,y,z)$.
\end{itemize}
Contracting the color indices with $\epsilon^{a b c}$ yields a gauge invariant operator, in other words a baryon with a red, a green and a blue quark. Suitable choices for $\Gamma$ and their corresponding quantum numbers are collected in Table~\ref{TAB002}, where
\begin{itemize}
\item $\psi^{(1)} \psi^{(2)} = ud - du$, if $I = 0$ and $S = 0$,

\item $\psi^{(1)} \psi^{(2)} \in \{ uu \, , \, dd \, , \, ud + du \}$, if $I = 1$ and $S = 0$,

\item $\psi^{(1)} \psi^{(2)} \in \{ us \, , \, ds \}$, if $I = 1/2$ and $S = -1$,

\item $\psi^{(1)} \psi^{(2)} = ss$, if $I = 0$ and $S = -2$.
\end{itemize}
For numerical results, plots and their discussion cf.\ \cite{Wagner:2011fs}.

\begin{table}[htb]
\begin{center}
\begin{tabular}{|c||c|c||c|c|c||c|c|c||c|c|c|}
\hline
 & & & & & & & & & & & \vspace{-0.40cm} \\
$\Gamma$ & $j^P$ & $J$ & $I$ & $S$ & name & $I$ & $S$ & name & $I$ & $S$ & name \\
 & & & & & & & & & & & \vspace{-0.40cm} \\
\hline
 & & & & & & & & & & & \vspace{-0.40cm} \\
$\gamma_5$ & $0^+$ & $1/2$ & $0$ & $0$ & $\Lambda_b$ & $1/2$ & $-1$ & $\Xi_b$ & X & X & X \\
$\gamma_0 \gamma_5 $ & $0^+$ & $1/2$ & $0$ & $0$ & $\Lambda_b$ & $1/2$ & $-1$ & $\Xi_b$ & X & X & X \\
$1$ & $0^-$ & $1/2$ & $0$ & $0$ & & $1/2$ & $-1$ & & X & X & X \\
$\gamma_0$ & $0^-$ & $1/2$ & $1$ & $0$ & & $1/2$ & $-1$ & & $0$ & $-2$ & \\
 & & & & & & & & & & & \vspace{-0.40cm} \\
\hline
 & & & & & & & & & & & \vspace{-0.40cm} \\
$\gamma_j$ & $1^+$ & $1/2$, $3/2$ & $1$ & $0$ & $\Sigma_b$, $\Sigma_b^\ast$ & $1/2$ & $-1$ & & $0$ & $-2$ & $\Omega_b$ \\
$\gamma_0 \gamma_j$ & $1^+$ & $1/2$, $3/2$ & $1$ & $0$ & $\Sigma_b$, $\Sigma_b^\ast$ & $1/2$ & $-1$ & & $0$ & $-2$ & $\Omega_b$ \\
$\gamma_j \gamma_5$ & $1^-$ & $1/2$, $3/2$ & $0$ & $0$ & & $1/2$ & $-1$ & & X & X & X \\
$\gamma_0 \gamma_j \gamma_5$ & $1^-$ & $1/2$, $3/2$ & $1$ & $0$ & & $1/2$ & $-1$ & & $0$ & $-2$ & \vspace{-0.40cm} \\
 & & & & & & & & & & & \\
\hline
\end{tabular}
\caption{\label{TAB002}$b$ baryon creation operators and their quantum numbers ($j^\mathcal{P}$: angular momentum of the light quarks and parity; $J$: total angular momentum; $I$: isospin; $S$: strangeness; name: name of the corresponding $b$ baryon(s) in \cite{PDG}); operators marked with ``X'' are identically zero, i.e.\ do not exist.}
\end{center}
\end{table}

For more lattice literature on $b$ baryons cf.\ \cite{Michael:1998sg,Detmold:2007wk,Burch:2008qx,Detmold:2008ww,Lin:2009rx,Lin:2010wb}.


\section*{Acknowledgments}

We thank HGS-HIRe for FAIR, in particular Gerhard Burau, for organizing the fruitful and nice ``HGS-HIRe Lecture Week on Hadron Physics'' in Laubach.

M.W.\ thanks Gerhard Burau for the invitation to the lecture week. M.W.\ thanks Chris Michael for countless helpful discussions on lattice hadron spectroscopy. M.W.\ acknowledges support by the Emmy Noether Programme of the DFG (German Research Foundation), grant WA 3000/1-1.

This work was supported in part by the Helmholtz International Center for FAIR (HIC for FAIR) within the framework of the LOEWE program launched by the State of Hesse and the Hemholtz Graduate School for Hadron and Ion Research (HGS-HIRe for FAIR).



\end{document}